\documentclass[11pt,preprint]{emulateapj}

\shorttitle{HCN (3-2) emission in the envelope of W Hya}
\shortauthors{Muller et al. 2008}
\begin{document}

\title{Distribution and kinematics of the HCN(3-2) emission down to the innermost region
in the envelope of the O-rich star W~Hya}

\author{S\'ebastien Muller \altaffilmark{1}, Dinh-V-Trung \altaffilmark{1,2},
Jin-Hua He \altaffilmark{1,3} \& Jeremy Lim \altaffilmark{1}}

\altaffiltext{1}{Academia Sinica, Institute of Astronomy and Astrophysics (ASIAA), P.O. Box 23-141, Taipei 106, Taiwan}
\altaffiltext{2}{On leave from Institute of Physics, Vietnamese Academy of Science \& Technology, 10, Daotan, BaDinh, Hanoi, Vietnam}
\altaffiltext{3}{New address: National Astronomical Observatories/Yunnan Observatory, Chinese Academy of Sciences, PO Box 110, Kunming, Yunnan Province 650011, China}

\email{muller, trung, jhhe, jlim @asiaa.sinica.edu.tw}

\begin{abstract}
We report high angular resolution observations of the HCN (3-2) line emission in the circumstellar envelope
of the O-rich star W~Hya with the Submillimeter Array. The proximity of this star allows us to image its
molecular envelope with a spatial resolution of just $\sim 40$ AU, corresponding to about 10 times the stellar
diameter. We resolve the HCN (3-2) emission and find that it is centrally peaked and has a roughly spherically
symmetrical distribution. This shows that HCN is formed in the innermost region of the envelope (within
$\sim 10$ stellar radii), which is consistent with predictions from pulsation-driven shock chemistry models,
and rules out the scenario in which HCN forms through photochemical reactions in the outer envelope. Our model
suggests that the envelope decreases steeply in temperature and increases smoothly in velocity with radius,
inconsistent with the standard model for mass-loss driven by radiative pressure on dust grains. We detect a
velocity gradient of $\sim 5$ km~s$^{-1}$ in the NW--SE direction over the central 40 AU. This velocity
gradient is reminescent of that seen in OH maser lines, and could be caused by the rotation of the envelope
or by a weak bipolar outflow.
\end{abstract}

\keywords{circumstellar matter --- radio lines: stars --- stars: individual (W~Hya) ---
stars: late-type --- stars: winds, outflows}

\section{Introduction}

The detection at radio frequencies of various molecules in the circumstellar envelopes of evolved stars
has considerably improved our understanding of circumstellar chemistry. This is particularly true for
the prototype of carbon stars, IRC+10216, on which much attention has been focused. Several dozens of
molecules have been detected in its circumstellar envelope (e.g., \citealt{cer00}), reflecting its
complex and rich chemistry. The inventory of molecules together with high angular resolution images of
their spatial distribution have led to the construction of sophisticated chemical models for the
circumstellar envelope of IRC+10216 (e.g., \citealt{gla87,mil94}) that now form the basis of our
understanding of the chemistry in C-rich envelopes.

The chemistry in O-rich envelopes, on the other hand, was not expected to be as rich because equilibrium
chemistry calculations indicate that nearly all the carbon should be locked into CO and nitrogen into
N$_2$ (\citealt{tsu73}). Indeed, before 1985, the inventory of molecules detected around oxygen stars
was sparse and did not include carbon-bearing molecules, except of CO. \cite{deg85} reported the first
detection of HCN towards O-rich stars. The list of HCN detections was later extended by \cite{jew86},
\cite{lin88}, \cite{ner89}, \cite{olo98} and \cite{bie00}. Recently, \cite{ziu07} also emphasized the
chemical complexity in the envelope of the O-rich supergiant star VY~CMa, following the detection of
various molecules such as HCO$^+$, CS, NaCl, PN and SiS.

Two competing models have been proposed to account for the formation of HCN in O-rich envelopes:
photochemical reactions in the outer envelope (\citealt{cha95}), or gas-phase non-equilibrium
chemical reactions in the inner region close to the stellar photosphere (within few tens of AU
from the star) due to shocks driven by stellar pulsation (\citealt{dua99,dua00}). The two models
predict very different HCN spatial distribution: in a hollow shell-like structure of radius 200
to 1000 AU depending on the mass loss for photochemical reactions, or a centrally concentrated
distribution extending up to the photodissociation radius of HCN for the shock-driven chemical
reactions.

Current observational data seem to favor pulsation-driven shock chemistry formation of HCN in O-rich
stars. \cite{bie00} conducted single-dish observations toward a sample of 16 O-rich stars in the
HCN (3-2) and (4-3) lines. The detection of these high density tracers is inconsistent with
photochemical production of HCN in the outer envelope. The detection of the HCN (3-2) (0,1$^{1c}$,0)
and (8-7) transitions toward $\chi$ Cyg gave further support for the formation of HCN in the innermost
region of the envelope, within $\sim 20$ stellar radii (i.e. $\sim 30$ AU), based on excitation
arguments (\citealt{dua00}). \cite{mar05} reported the first interferometric observations of HCN in
the circumstellar envelopes of the O-rich stars IK~Tau and TX~Cam. In both stars, the HCN emission
appears to be largely concentrated within $\sim 800$ AU, as predicted by shock chemistry models.



If HCN indeed forms close to the stellar photosphere, it should be an excellent tracer of the inner
envelope, and perhaps even of the wind acceleration zone where, in the standard model, dust particles
form and radiation pressure on dust becomes most effective. Maser (e.g., H$_2$O or OH)
emission in the inner envelope can be observed with very long baseline interferometry, but traces
only a limited radial range. On the other hand, there are so far only few reports of high angular
resolution observations of other thermal lines in the inner envelope, such as SiO (2-1) $\nu = 0$
(\citealt{luc92,sah93,sch04}), none of them with angular resolution higher than $\sim 100$ AU.


In this letter, we report very high spatial resolution ($\sim 40$ AU) observations of the HCN (3-2)
emission toward W~Hya with the Submillimeter Array. 
W~Hya is one of the closest O-rich stars at a distance of 78 pc (based on Hipparcos data, \citealt{per97},
revised by \citealt{kna03}), and shows the brightest HCN (3-2) and (4-3) emission amongst the sample
of O-rich stars observed by \cite{bie00}.
It is therefore an excellent target in which to investigate the distribution and kinematics of HCN in the
circumstellar envelope of an O-rich star.


\section{Observations} \label{obs}

We observed W~Hya on 2008 April 13 and 15 with the Submillimeter Array
\footnote{The Submillimeter Array is a joint project between the Smithsonian Astrophysical Observatory
and the Academia Sinica Institute of Astronomy and Astrophysics and is funded by the Smithsonian
Institution and the Academia Sinica.}
(SMA) in its very extended
configuration. All eight antennas were operating on the first day of our observations, and six on
the second day. The zenith atmospheric opacity was $\sim 0.2$ at 225 GHz during both nights,
resulting in system temperatures that changed between 200 and 600 K depending on source elevation.

The heterodyne SIS receivers were tuned to the frequency of the HCN (3-2) transition at 265.886 GHz in
the lower side band. The correlator was configured to give a spectral resolution of 0.8125 MHz, which
corresponds to a velocity resolution of $\sim 0.9$ km~s$^{-1}$. The line-free channels in both the
upper and lower side bands, spanning a total bandwidth of $\sim 3.8$ GHz at 1.1 mm, were used to make
a map of the continuum. The bandpass of the individual antennas was derived from the bright quasar
3C273. Flux calibration was derived by observing Titan and Callisto. The quasars 1334$-$127 and
1313$-$333, located within $16\degr$ of W~Hya, were observed every 15 min for complex gain calibration.
Data reduction was done separately for both tracks using MIR/IDL. The calibrated visibilities were
written out in FITS format, and then converted into GILDAS format for imaging purpose using MAPPING. 

The projected baselines ranged from 22 to 510 m. Natural weighting of the visibilities yielded a
synthesized beam of $0.55\arcsec \times 0.40\arcsec$ at a position angle of $26\degr$. We used the
CLARK algorithm to deconvolve the image. For the HCN (3-2) channel map, we estimate the $1\sigma$ rms
noise level to be 15 mJy/beam in each 0.9 km~s$^{-1}$ channels. For the 1.1 mm continuum, the noise
level is 3 mJy/beam. The primary beam of the SMA antennas spans a FWHM of $46.5\arcsec$ at 266 GHz.


\section{Results} \label{results}

\begin{figure} \includegraphics[width=8cm]{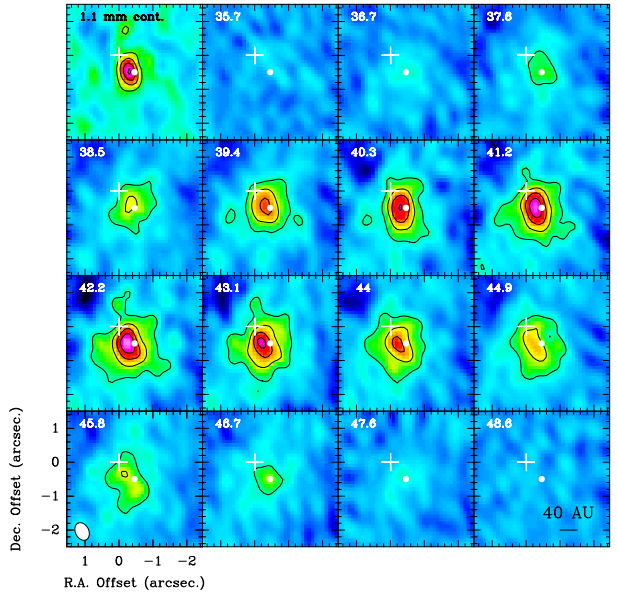}
\caption{1.1 mm continuum emission ({\em top left corner box}) and HCN (3-2) channel maps of W~Hya as observed
with the SMA very extended configuration. Contours are every 25 mJy/beam for the continuum map,  and every
0.75 Jy/beam ($5\sigma$), corresponding to 60 K for the synthesized beam of $0.55\arcsec \times 0.40\arcsec$
(P.A.$ = 26\degr$), for the channel maps. The white cross indicates the position of the phase center, set at the
stellar position from the Hipparcos catalog in the year 2000
(R.A.$_{J2000} = 13^{\rm h} 49^{\rm m} 01\fs998$ and
Dec.$_{J2000} = -28 \degr 22 \arcmin 03 \farcs 49$). The white dot gives the expected position of
the star at the date of our observations in 2008, due to its proper motion. The small offset ($\lesssim 0.15\arcsec$)
between this position and the peak emission may be due to baseline errors.}
\label{map} \end{figure}

The HCN (3-2) channel maps are shown in Figure \ref{map} together with the 1.1 mm continuum map.
The molecular emission, which is clearly resolved, is centrally peaked and has a roughly spherically
symmetrical distribution. The continuum emission is unresolved and has a total flux density of
$270 \pm 15$ mJy at 1.1 mm, consistent with the value of $280 \pm 30$ mJy obtained by \cite{vanvee95}
with the James Clerk Maxwell Telescope. We estimate a stellar blackbody contribution of 225 mJy based
on the stellar parameters given in Table 1, and which therefore dominates the 1.1 mm continuum emission.

W~Hya is known to have a very extended dusty envelope with a radius of $\sim 30\arcmin$ (\citealt{haw90}).
The photodissociation radius of HCN, however, is expected to be much smaller, of order of $1\arcsec$,
based on computations by \cite{olo98} (see their Eq.2). From a Gaussian fit of the visibilities, we find
that the HCN (3-2) emitting region has a radius at FWHM of about $0.6\arcsec$, comparable to the size
measured by \cite{luc92} for the SiO (2-1) emission. Comparing our data with that from single-dish
observations by \cite{bie00}, we estimate that we recover $\ga 70$\% of the total line intensity within
$28\arcsec$ of the star. A spectrum of the HCN (3-2) line, extracted at the peak position of the emission,
is shown in Figure \ref{spec}. The peak brightness temperature of the line is $\sim 300$ K. The line profile
is clearly non-symmetric with respect to the systemic velocity of $\sim 40$ km~s$^{-1}$ (\citealt{cer97}).

A visual inspection of the channel maps reveals a weak velocity gradient in the NW to SE direction.
A fit of the visibilities yields a velocity gradient of $\sim 5$ km~s$^{-1}$ at a position angle of
$125\degr$ over the $0.5\arcsec$ central region. Position-velocity diagrams along this direction,
as well as on a perpendicular axis, are shown in Figure \ref{pv}.




\section{Modeling} \label{model}

\begin{figure} \includegraphics[width=8cm]{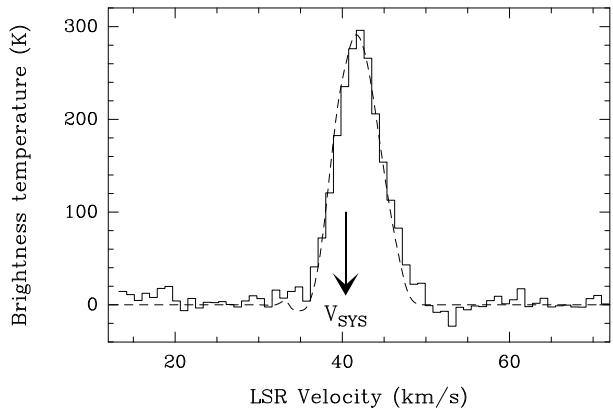}
\caption{Brightness temperature at the peak 
of the HCN (3-2) emission as a function of velocity, with model prediction overlaid ({\em dashed line}).
Note that the blue-shifted part of the spectrum is strongly affected by self-absorption.}
\label{spec} \end{figure}

\begin{figure} \includegraphics[width=8cm]{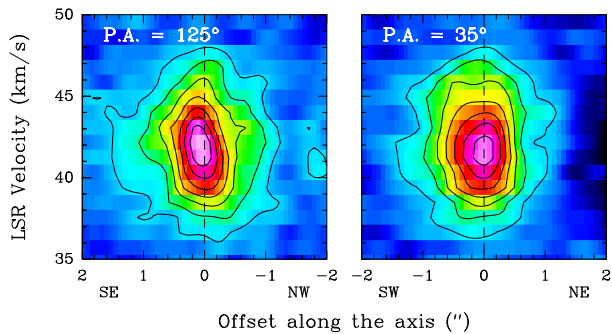}
\caption{Position-velocity diagrams along two perpendicular axes. Note the small velocity gradient along
the SE-NW direction.}
\label{pv} \end{figure}

\begin{figure} \includegraphics[width=7cm]{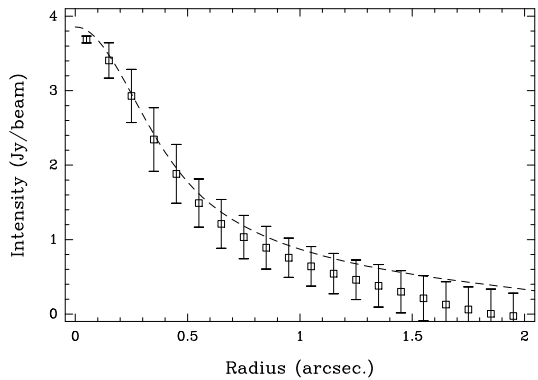}
\caption{Azimuthal average of the HCN (3-2) emission as a function of radius at V$_{\rm LSR}$ = 42.1 km~s$^{-1}$,
corresponding to the peak of the line profile in Figure \ref{spec}.
Our SMA data are indicated by open squares with error bars overlaid. The model prediction is shown in dashed line.}
\label{radial} \end{figure}

The peak brightness temperature, line profile and radial distribution of the HCN (3-2) gas provide very
useful constraints on the wind velocity and the physical conditions of the gas in the vicinity of W~Hya.
We used the molecular excitation and radiative transfer code of \cite{din00} to interpret our observations
and other relevant data. We assumed a spherical symmetry and did not attempt to reproduce the weak velocity
gradient. The model parameters are summarized in Table 1 and reproduce remarkably well our data
(see Fig.\ref{spec} \& \ref{radial}).

In the inner region of the envelope, the HCN molecules are excited by collisions with H$_2$ and by the
absorption of stellar photons at 3 $\mu$m (to $\nu_3 = 1$ vibrational state), 7 $\mu$m (to 02$^0$0
vibrational state), and 14 $\mu$m (to 01$^1$0 vibrational state). In the case of W~Hya, where little
absorption by dust is expected, the pumping route through the $\nu_3$ = 1 and 01$^1$0 states are the
most important because of the strong stellar radiation field at the corresponding wavelengths and
large transition rates. In our code, we take into account the vibrational ground state, (01$^1$0) and
$\nu_3 = 1$ vibrational states, and include explicitly all hyperfine levels up to $J=15$. In W~Hya, we
find that the $\nu_3 = 1$ absorption band of HCN falls within the deep H$_2$O absorption band
(\citealt{jus04}). The water molecules responsible for this absorption band are hot and expected to be
very close to the central star, presumably interior to the HCN emitting region. Thus, for the sake of
simplicity, we reduced the stellar radiation field by 50\% at 3 $\mu$m.

The mass loss rate of W~Hya is still quite uncertain. \cite{zub00} derived a mass loss rate of about
10$^{-6}$ M$_\odot$~yr$^{-1}$ (scaled to the revised distance of 78 pc), while \cite{jus05} favored
a much lower rate of $2-3 \times 10^{-7}$ M$_\odot$~yr$^{-1}$. We adopt here a value of $5 \times
10^{-7}$ M$_\odot$~yr$^{-1}$. The abundance of HCN relative to molecular hydrogen predicted from
non-equilibrium chemical models (\cite{dua99,dua00}) may vary between a few times 10$^{-6}$ and a
few times 10$^{-5}$, depending on the shock velocity. We adopt a representative constant value of 10$^{-6}$.
We note, however, that this assumption leads to a very slight overestimate of the HCN (3-2) intensity in the
outer part of the HCN envelope (see Fig.\ref{radial}). Most likely, the abundance of HCN is gradually decreasing
with radius due to the photodissociation by external radiation field.
We also assumed a local turbulent velocity of 1 km~s$^{-1}$ in the envelope.

We tried various expansion velocity laws and found that the relation V(r) $\propto$ log(r), starting
from 2 km~s$^{-1}$ at an inner radius of $\sim 10^{14}$ cm and reaching 7 km~s$^{-1}$ at a radius of
$10^{16}$ cm, produces a satisfactory fit to the HCN (3-2) line profile, and is consistent with OH
maser observations of \cite{szy98}. Other faster rising velocity laws, such as those prescribed by
\cite{deg90} or \cite{zub00}, do not reproduce the rounded-top and asymmetric profile of HCN (3-2)
as seen by the SMA (they actually produce double-peak profiles).

The rounded-top line profile indicates that the HCN (3-2) is optically thick. In addition, we find that
the excitation temperature of the HCN (3-2) transition follows closely the kinetic temperature of the gas.
Unless the wind is highly clumpy, resulting in small filling factor in the beam, the brightness temperature
of the HCN (3-2) emission, resolved by the SMA, is therefore directly related to the temperature profile
in the inner region. The very high temperature profile used by \cite{deg90} or the one designated as GS in
Figure 3 of \cite{zub00} are clearly inconsistent with our data. The shallower temperature profile derived
by \cite{zub00} also produces too high a brightness temperature for the HCN (3-2) line because the kinetic
temperature is above 400 K within an inner region of 50 AU in diameter, which is comparable to the SMA
synthesized beam of $\sim 0.5\arcsec$. In our model, we need to use a steep temperature profile T(r)
$\propto$ r$^{-1}$, starting with T$_0 = 650$ K at the base of the envelope. The predicted line profile is
compared with that observed in Figure \ref{spec}. Our model also provides a good match to the single-dish
profiles of the HCN (3-2) and (4-3) lines observed by \cite{bie00}.

We note that for the HCN (3-2) line,
the resulting profile is asymmetric and does not peak at the systemic velocity. That is due to the strong
self-absorption arising from the combination of steep temperature profile and slow acceleration in the
inner envelope. This effect can also explain the difference in the inferred systemic velocity from Gaussian
fitting of optically thick lines such as HCN (4-3), SiO(5-4) and (8-7) (\citealt{bie00}), with respect
to other lines such as SiO(2-1) and CO lines (\citealt{cer97}).

\begin{table}
\begin{tabular}{lll}
\multicolumn{3}{c}{TABLE 1: MODEL PARAMETERS} \\
\hline
\hline
\multicolumn{3}{c}{STAR} \\
Distance & D = 78 pc & (1) \\
Systemic velocity (LSR) & V$_{\rm SYS} = 40.4$ km~s$^{-1}$ & (2) \\
Effective temperature & T$_{\rm eff,\star} = 2500$ K & (3) \\
Stellar radius & R$_\star$ = $2.73 \times 10^{13}$ cm & (4) \\
\multicolumn{3}{c}{ENVELOPE} \\
Inner radius & R$_{\rm in} = 1 \times 10^{14}$ cm &  \\
Outer radius & R$_{\rm out} = 3 \times 10^{15}$ cm & \\
Mass-loss rate & $\dot{\rm M} = 5 \times 10^{-7}$ M$_\odot$ yr$^{-1}$ & \\
Temperature profile & T(r) = 650 K (r/R$_{\rm in}$)$^{-1}$ & \\
Launching velocity & V$_0 = 2$ km~s$^{-1}$ & (5) \\
Terminal velocity & V$_{\infty} = 7$ km~s$^{-1}$ & (4) \\
Velocity law (r $<$ R$_{\rm out}$) & V(r) =  $\frac{ \rm (V_{\infty}-V_0) log_{10} (r/R_{in}) }{\rm log_{10} (10^{16}{~\rm cm} /R_{in}) } +V_0 $ & \\
Abundance of HCN & [HCN]/[H$_2$] = 10$^{-6}$ & \\
Local turbulent velocity & $\sigma_{\rm turb} = 1$ km~s$^{-1}$ & \\
%
%
\hline
\end{tabular}
\tablerefs{ (1) \cite{kna03}; (2) \cite{cer97}; (3) \cite{han95}; (4) \cite{jus05}; (5) \cite{miy94}.}
\end{table}

\section{Discussion and perspectives} \label{discuss}

In the model proposed by \cite{cha95}, HCN is formed in the outer envelope as a result of photochemical
reactions initiated from methane. The distribution of HCN is predicted to peak at a radius $\ge 200$ AU
for a mass-loss rate $\ge 10^{-7}$ M$_\odot$~yr$^{-1}$. Our SMA observations, however, clearly reveal
that in the case of W~Hya, HCN is present, and thus should form, much closer to the stellar photosphere,
within 20 AU. This is consistent with the shock chemistry model developed by \cite{dua99} and \cite{dua00}.
In this model, stellar pulsations induce strong and periodic shocks in a narrow region above the photosphere.
Thermal equilibrium (TE) abundances are assumed as initial conditions in the photosphere, and the evolution
of chemical abundances is investigated in the post-shock region. As a net result of shock chemistry, several
carbon-bearing species, such as HCN, CS and CO$_2$, are produced in significant amounts. Noticeably, the
abundance of HCN can be increased by about 5 orders of magnitude with respect to its former TE value. The
route to HCN is through the reaction of CN with H$_2$, which is sensitive to temperature and becomes very
efficient in the post-shock layer. Since HCN is chemically stable and does not participate in the formation
of dust grains (unlike, e.g., SiO) in the inner envelope, it further remains unaltered as it travels through
the envelope, until it reaches the photodissociation region of the outer wind. As a result, HCN is a good
tracer of the kinematics in the envelope.

The requirement of a smoothly increasing velocity with radius to reproduce the HCN (3-2) line profile for
W~Hya and maser observations (\citealt{szy98}) are inconsistent with a fast acceleration of the wind within
$\la 10$ stellar radii, as predicted by models of mass-loss driven by radiative pressure on dust grains
(e.g., \citealt{kwo75,gol76}). Observations in other lines tracing different radii also called for a slow
wind acceleration, which could be explained if dust grain formation occurs over large extent up to several
hundreds of AU (\citealt{luc92}), and/or if grain properties change through the envelope (\citealt{cha86}).
Alternatively, the slow acceleration could result from the inefficiency of dust drag when the mass loss is
small (see, e.g., \citealt{bow88,win00}).


The velocity gradient along the SE--NW axis (see Fig.\ref{pv}) is about 5 km~s$^{-1}$ over $0.5\arcsec$.
MERLIN observations of OH maser lines by \cite{szy98} reveal a remarkably similar trend in amplitude
and position angle. Interestingly, \cite{lat97} measured the visible diameter of W~Hya along two orthogonal
directions and found a significant asymmetry of 20\%, with elongation in a direction comparable to that of
the velocity gradient seen in the envelope. It is not clear, however, whether this apparent elongation is
due to some oblateness of the star or to a nonsymmetric brightness distribution over the stellar disk.
The velocity gradient that we observe in the inner envelope of W~Hya could be either due to the rotation of the
envelope or to a weak bipolar outflow.


The next jump in angular resolution achievable with the Atacama Large Millimeter Array will allow us to
probe the envelope even deeper, possibly down to the dust formation zone.
Such observations could directly test the validity of shock chemistry models.
In particular, very high angular resolution observations at different phases of the stellar variablity
could directly show the effects of the pulsations on the 
formation of dust and molecules, and possible modulations of the mass loss (see, e.g., \citealt{dia03}).







\end{document}